\documentstyle[aps,multicol,psfig]{revtex}

\begin{document}
\draft 

\title{Reflectionless tunneling in ballistic normal-metal--superconductor 
junctions}

\author{M. Schechter, Y. Imry, and Y. Levinson}

\address{Dept. of Condensed Matter Physics, The Weizmann
Institute of Science, Rehovot 76100, Israel}


\maketitle
   
\vspace{0.1cm}

\begin{abstract} 
We investigate the phenomenon of reflectionless tunneling in ballistic 
normal-metal--superconductor 
(NS) structures using a semiclassical formalism.  It is shown that
applied magnetic field and superconducting phase difference both
impair the constructive interference leading to this effect, but in a
qualitatively different way.  This is manifested both in the
conductance and in the shot noise properties of the system considered.
Unlike diffusive systems, the features of the conductance are sharp
and enable fine spatial control of the current, as well as 
single-channel manipulations.  We discuss the possibility of conducting
experiments in ballistic semiconductor-superconductor structures with
smooth interfaces and some of the phenomena specific to such structures 
that could be measured. 
A general criterion for the barrier at NS
interfaces, though large, to be effectively transparent to pair
current is obtained.
\end{abstract}    
\vspace{0.8cm}   

\begin{multicols}{2}
\section{Introduction} 

One of the most interesting phenomena in hybrid diffusive 
normal-metal--superconductor structures is reflectionless tunneling.  
This
phenomenon manifests itself as a zero bias peak in the differential
conductance of a diffusive normal metal slab connected to a
superconductor via a tunnel barrier with low transmission probability
$\Gamma$ \cite{KKG+91,NKH92}. van Wees {\it et al.}
\cite{WVMK92} used a path integral picture to suggest and explain the effect 
of reflectionless tunneling.  They show that the enhanced conductance
at zero-bias is due to electron-hole coherence in trajectories that
due to the disorder in the normal metal hit the barrier at the 
normal-metal--superconductor (NS) interface many times. This results 
in the barrier being effectively transparent to pair current.

In this paper we show that the phenomenon of reflectionless tunneling
exists also in ballistic systems, the requirement being the existence
of multiple reflections from the NS interface due to the geometry of
the structure.  As in diffusive systems, we find an enhanced NS
conductance for zero-bias and zero magnetic field.  We show that the 
magnetic field ($H$), finite energy and voltage, and superconducting phase
difference ($\Phi_s$) impair the constructive interference leading to
the enhanced NS conductance, but applying the superconducting phase
difference has qualitatively different consequences than applying a 
finite magnetic field or voltage.

We show that the ballistic nature of the system gives rise to
pronounced and delicate features, which are not averaged over as in
the case of diffusive systems.  This results in new measurable
phenomena, such as sharp {\it peaks} in the NS conductance as new
channels open, and quasiperiodicity of the conductance as a function of
magnetic field. We also demonstrate the possibility, specific to
ballistic systems, to conduct detailed manipulations such as
extracting out a single channel from a normal metal (semiconductor)
waveguide or extracting the current at a given position along the
waveguide.

The ballistic regime in semiconductor-superconductor hybrid structures
was investigated recently experimentally
\cite{NATA92,TAN95,TAN95b,DHW+95,MHW+97,MWK+97,HWKB99}. 
Unlike the case in normal-metal--superconductor structures, where sharp
boundaries are made that enable specular reflection at the NS
interface \cite{BKW83,BGKW85}, in semiconductor-superconductor
interfaces specular reflection is sacrificed for the purpose of
lowering the barrier at the interface, thus increasing the Andreev
reflection probability.  We here raise the possibility to conduct
experiments in ballistic semiconductor-superconductor structures with
a sharp interface and a long elastic mean free path. Though indeed the
transmission probability of the barrier would then be small, 
the electron-hole coherence over long trajectories 
results in a large Andreev reflection probability, as we show below.
Thus, one can have strong proximity while preserving the ballistic
nature of the system.  Other systems which seem favorable for the 
realization of ballistic NS structures with specular reflection at the
interface are the recently investigated organic molecular crystals
\cite{SKHB00,SKB00}.  In these systems the NS transition could be
realized by applying a space dependent gate voltage.

The paper is arranged as follows: In Sec.~\ref{secNtrajectory} we
introduce the formalism and the structure we consider, obtain the
expressions for the three-terminal conductances in terms of
$R_{he}(N)$, the Andreev reflection probability of a trajectory that
hits the interface $N$ times, and calculate this probability for zero
magnetic field.  In Sec.~\ref{secopening} we show that for a short
slab the NS conductance has sharp {\it peaks} as channels open.  In
Secs.~\ref{secmagnetic} and \ref{secSNS} we calculate $R_{he}(N)$ and the
linear conductances as function of $H$ (\ref{secmagnetic}) and
$\Phi_s$ in a similar SNS structure (\ref{secSNS}).  In
Sec.~\ref{secnoise} we calculate the shot noise in both structures, as
a function of $H$ and $\Phi_s$.  In Sec.~\ref{secdiffusive} we consider
diffusive systems and demonstrate the connection between the effect
of reflectionless tunneling in diffusive and ballistic systems.
Throughout the paper we consider zero temperature and use the model in
which the superconducting order parameter $\Delta$ is constant in the
superconductor and zero in the normal metal.

\section{Conductance of a long normal slab attached to a superconductor}
\label{secNtrajectory}

\subsection{Model}

\begin{figure} 
 	\narrowtext 
	\centerline{\psfig{figure=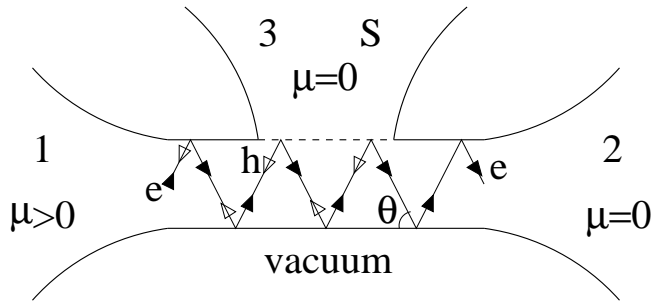,width=3.375in}}
	\vspace{1 cm}
	\caption{Vacuum--ballistic normal-metal--superconductor junction with 
	a barrier at the NS interface. Each 
	time the particle hits the NS interface it can be reflected either  
	normally or in an Andreev process.$1$ and $2$ are normal 
	reservoirs and $3$ is a superconducting reservoir. 
	$\theta$ is the angle of incidence. Filled (empty) arrows 
	designate electrons (holes). 
	In Sec.~\protect\ref{secSNS} a similar structure is considered, 
	with a second superconductor attached to the other (lower) side 
	of the slab, so the structure has up-down symmetry.}
	\label{newvns}
\end{figure}

We consider a ballistic normal-metal or semiconductor slab between two
normal reservoirs. The slab is separated by an infinite barrier from a
region denoted as vacuum, except in a region of length $L$, at which a
superconductor is attached to the slab (Fig.~\ref{newvns}). At the NS
interface the barrier is finite, with transmission probability
$\Gamma$.  The opening of the normal slab to the two normal reservoirs
is taken to be adiabatic and the length of the slab between the
reservoirs and the NS interface to be long enough such that channels
are formed with homogeneous distribution in the transverse direction
\cite{SL90}.  We also assume that the change from infinite barrier to
finite barrier of transmission $\Gamma$ at the end points of the NS
interface is not abrupt, but smeared over a length $s$ such that
$\lambda_F \ll s \ll W$, where $W$ is the width of the slab in the
direction perpendicular to the interface.  In this way the change is
adiabatic but the smearing can be neglected in our calculations.  We
denote this structure as a vacuum--normal-metal--superconductor (VNS)
structure, as opposed to a similar structure with another
superconductor attached symmetrically to the other side of the slab,
which will be an denoted SNS structure.

The superconductor is connected to a third reservoir except when
explicitly mentioned otherwise.  We consider the case where the
electrochemical potentials of the right and superconducting reservoirs
are equal, and the left reservoir is biased by an infinitesimal
voltage, and calculate the three-terminal linear conductances of the
system.  Previous works concerning similar structures
\cite{KZSJ95,CRL96,ASRL96,BKZ+98} considered the NS interface either
as fully transparent or concentrated on effects of channel mixing due
to the roughness of the barrier when it exists. We consider the NS
interfaces to have a smooth barrier, so that normal reflection is
specular and the Andreev reflected hole retraces the electron's
trajectory.  We assume specular reflection from the VN interface as
well.

Our model is two dimensional. While assuming $\lambda_F \ll W$,
therefore having many channels, we assume for simplicity that the
thickness of the slab (the third dimension) is small, having one
transverse mode in this direction.  The generalization of our
treatment to thicker slabs is trivial. 
 
We use a semiclassical formalism and consider the propagation of
electrons in each channel to be described by their classical
deterministic trajectory \cite{WVMK92,Cht01}.  For each channel $j$ we
define $k_{j\|} = \sqrt{2 m E_F/\hbar^2 - j^2 \pi^2/W^2}$ and
calculate the angle $\theta_j = \tan^{-1}[j \pi/(k_{j\|} W)]$ between
the classical trajectory of an electron in this channel and the NS
interface.  We consider an electron entering the normal slab from the
left reservoir, approaching the region of the slab with the NS
interface (``NS region'') at a given distance from the NS interface
and angle $\theta_j$ with respect to it.  If the electron is only
normally reflected from the NS interface, it follows a certain
trajectory in the slab until exiting it to the right reservoir after
hitting the NS interface $N$ times (``$N$ trajectory'').  Due to the
finite Andreev reflection amplitude at each point it hits the NS
interface, the electron has a probability $R_{he}(N)$ to be reflected
as a hole to the left reservoir.  In this model, due to the interfaces
being parallel and smooth, there is zero probability for an electron
to be reflected back to the left reservoir or to be transmitted as a
hole to the right reservoir. Therefore, $R_{he}(N) + T_{ee}(N) = 1$,
where $T_{ee}(N)$ is the probability of an electron coming from the
left reservoir to be transmitted as an electron to the right
reservoir.

\setlength{\unitlength}{3in} 

For each open channel in the slab the number of times a trajectory
hits the NS interface is either $N_j$ or $N_j+1$, where $N_j$ equals
the integer part of $L \tan{\theta_j}/(2 W)$, with $L$ being the
length of the NS interface.  The fraction of trajectories in channel
$j$ that hit the NS interface $N_j+1$ times is given by 
$p_j = L \tan{\theta_j}/(2 W) - N_j$.  The Andreev reflection probability 
of an electron in channel $j$ is then given by
\begin{equation}
R^j_{he} \equiv p_j R_{he}(N_j+1) + (1-p_j) R_{he}(N_j) \; . 
\label{Rlhe}
\end{equation} 

We define by $I_1, I_2$ and $I_3$ the currents emerging from the left 
terminal, right terminal, and superconducting terminal, respectively. 
Due to current conservation $I_1 = - I_2 - I_3$. 
We then define the NN, NS, and total linear conductances of the system as 

\begin{equation}
G_{21} \equiv - \lim_{V \rightarrow 0} \frac{I_2}{V}   = \frac{2 e^2}{h} 
\sum_j \Theta(k_{j\|}^2) (1 -  R^j_{he}) \; ,
\label{Glr}
\end{equation}

\begin{equation}
G_{31} \equiv - \lim_{V \rightarrow 0} \frac{I_3}{V}= \frac{4 e^2}{h} 
\sum_j \Theta(k_{j\|}^2) R^j_{he} \; ,
\label{Gls}
\end{equation}
and 

\begin{equation}
G_T  \equiv  \lim_{V \rightarrow 0} \frac{I_1}{V} =   G_{21} + G_{31} 
= \frac{2 e^2}{h} 
\sum_j \Theta(k_{j\|}^2) (1+R^j_{he}) \; ,
\label{GT}
\end{equation}
where $\Theta(x)$ is the Heaviside theta function. 

\subsection{Andreev reflection probability of an $N$ trajectory}
\label{NAndreev}

The calculation of the conductances is therefore reduced to the
calculation of $R_{he}(N) \equiv |r_{he}(N)|^2$ where $r_{he}(N)$ is
the corresponding amplitude.  For a single hit at the NS boundary, we
denote by $r_{he} \,\, (r_{eh})$ the amplitude for an electron (hole) to be
Andreev reflected and by $r_{ee} \,\, (r_{hh})$ the amplitude for an
electron (hole) to be normally reflected.  By dividing an $N$
trajectory to an $N-1$ trajectory and a $1$ trajectory, we obtain a
recursion formula

\begin{eqnarray} 
r_{he}(N) = r_{he} + r_{ee} r_{he}(N-1) r_{hh} + \nonumber \\
r_{ee} r_{he}(N-1) r_{eh} r_{he}(N-1) r_{hh} +... \nonumber \\
=r_{he} + \frac{r_{ee} r_{he}(N-1) r_{hh}}{1-r_{eh} r_{he}(N-1)} 
\; .
\label{rheN}  
\end{eqnarray} 

Using the relations (which are exact at $E_F$) 
$r_{ee}=r_{hh}^\ast$, 
$r_{eh}=r_{he}, $ and 
$|r_{eh}|^2+|r_{ee}|^2=1$, we assume, and then show by induction, 
that $r_{he}(N)$ is imaginary for all $N$ and can be written as  

\begin{equation} 
r_{he}(N)=i \frac{|r_{he}(N-1)| + |r_{he}|}{1 + |r_{he}| |r_{he}(N-1)|} \; .
\end{equation}  
The solution of this equation is given by 

\begin{equation} 
r_{he}(N) = i \tanh[N \tanh^{-1}(|r_{he}|)] \; .
\label{arctan} 
\end{equation} 
For a barrier with small transmission probability we find 

\begin{equation}
  R_{he}(N) \approx  \tanh^2(N r) \approx  \tanh^2(N \Gamma/2 ) \; , 
  \label{Rheapprox}
\end{equation} 
where we define $r \equiv |r_{he}| = \Gamma/(2-\Gamma)$. 

Using Eq. (\ref{arctan}) to obtain the values of $R_{he}(N)$ for all
the channel-dependent $N_j$ and $N_j+1$ in the conductance
formulas~[Eqs.~(\ref{Glr})-(\ref{GT})], we obtain the linear conductances
$G_{21}$, $G_{31}$, and $G_T$. 
In this paper we are interested in the case were $\Gamma \ll 1$ and, 
therefore, in some of the formulas, and in the qualitative discussions, 
we take this limit. 

Before considering further the conductances of the system, we would
like to dwell on the physical aspects of Eq.~(\ref{Rheapprox}). This
formula reflects the essence of the physics behind ``reflectionless
tunneling.''  It states that electrons in trajectories that hit the
NS interface $N \gg 1/\Gamma$ times are Andreev reflected with
probability close to unity, even though $\Gamma \ll 1$, thus making a
barrier having a low transmission coefficient effectively transparent
to pair current.  
\begin{figure} 
	\centerline{\psfig{figure=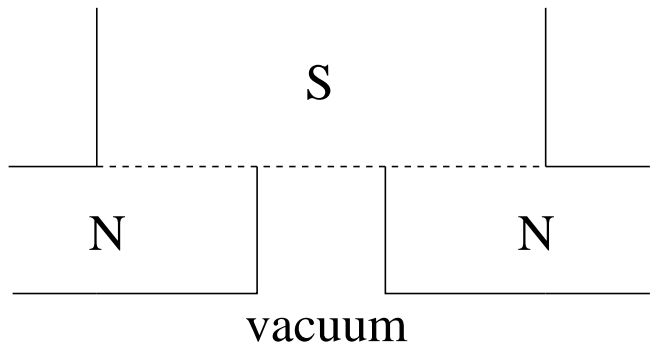,width=3.375in}}
	\vspace{0.3cm}
	\caption{Vacuum--ballistic normal-metal--superconductor junction where 
	a part of the normal metal is removed (only the relevant region 
	is shown).}
	\label{fignin}
\end{figure}
This is a result of electron-hole coherence in the
normal metal. For an incoming electron, the different paths resulting
in a hole returning to the reservoir interfere constructively, while
the different paths resulting in an electron transmitted to the right
reservoir interfere destructively.  The constructive interference for
a returned hole competes with the small Andreev amplitude at each
encounter with the interface, which is proportional to $\Gamma$, and
therefore $R_{he}(N) \approx 1$ only for $N \gg \Gamma^{-1}$.  This
means that for channels in which $\tan{\theta} \gg 2W/(\Gamma L)$ the
barrier at the NS interface is not effective.  In fact, if one
considers a system in which the superconductor is floating and the
above condition is fulfilled for all the channels, one can show that
the current between the left and right reservoirs flows inside the
superconductor and a part from the middle of the normal slab can be
taken out (see Fig.~\ref{fignin}) without affecting the conductance of
the system.

Equations~(\ref{arctan}) and (\ref{Rheapprox}) are far more general than
the above model and hold in any case where an electron in the normal
metal can hit the NS interface more than once before electron-hole
coherence is lost.  This is true in various geometries in ballistic
systems, and also in diffusive systems, which are considered in
Sec.~\ref{secdiffusive}.  In all these cases, Eq.~(\ref{Rheapprox})
results in a criterion for the effectiveness of a barrier with small
transmission probability: {\it Consider a physical property which is
determined by a certain set of trajectories; the criterion for the
barrier at the NS interface not to be effective is that most of these
trajectories hit the interface more than $\Gamma^{-1}$ times before
electron-hole coherence is lost.}  In Sec.~\ref{secdiffusive} we will
show how this general criterion reduces, in diffusive systems, to the
known conditions for the barrier, though high ($\Gamma \ll 1$), not to
affect the conductance and the density of states of a diffusive NIS
junction.

\subsection{Comparison to an incoherent structure}
\label{pairinglength}
 
In order to show that Eq.~(\ref{Rheapprox}) is a result of
constructive interference, which is due to electron-hole coherence, we
compare our result to the case where, due to strong dephasing, there
is no electron-hole coherence, i.e., where the phase between two consecutive
hits of the interface is lost.  In this case the problem is reduced to
a random walk problem, with forward-backward asymmetry.  At each hit
at the NS interface the electron (hole) has a probability 
$\Gamma^2 \ll 1$ to be Andreev reflected, in which case the direction of
propagation is reversed and the probability to move forward is
$1-\Gamma^2$.  The size of the step is channel dependent and is given
by $d_j=2 W \cot{\theta_j}$, the distance between two consecutive
points a trajectory in channel $j$ hits the interface.  This gives a
mean free path of $l_j=d_j/\Gamma^2$.  For $L \ll l_j$ the Andreev
reflection probability is $R_{he}=L/l_j$, and for $L \gg l_j$ it is 
$1 - l_j/L$ (the probability for a transmitted electron is $l_j/L$).
Therefore $l_j$ is the saturation length, beyond which the Andreev
reflection probability is close to unity.  On the contrary, in the case of
coherent scattering, using Eq.~(\ref{Rheapprox}) and the relation 
$N_j = L/d_j$, we find that the saturation length is 
$d_j/\Gamma = l_j \Gamma$. Due to the scattering being coherent, it is 
smaller by $\Gamma$ compared to the noncoherent case.  Moreover, for short 
slabs, $L \ll l_j \Gamma$, $R_{he} \approx L^2/(4 l_j d_j)$, larger by 
$L/(4 d_j)$ than the noncoherent case. For long slabs 
($L \gg l_j \Gamma \;$) 
one obtains $R_{he} \approx 1 - \exp(-L/\sqrt{2 l_j d_j})$, and the
probability for a transmitted electron is exponentially small, and not
linear in $l_j/L$ as in the noncoherent case. The difference between
the two cases is most notable for slabs with intermediate lengths
between the two saturation lengths, $l_j \Gamma \ll L \ll l_j$, which
corresponds to $1/\Gamma \ll N \ll 1/\Gamma^2$. Without coherence
$R_{he} \ll 1$, and with coherence $R_{he} \approx 1$.

It is instructive to compare Eq.~(\ref{Rheapprox}) to a similar
system, in which the superconductor is not attached to the slab on its
side, but part of the slab itself, of length $L$, is
superconducting. In this case and assuming no barrier at the NS
interfaces, the Andreev reflection probability of an incoming electron
is given by $\tanh^2(\Delta L /2 \hbar v_F) \equiv \tanh^2(L/2 \xi_s)$
\cite{MB94}.  Here it is $\xi_s$, the ballistic superconducting
coherence length, which is the length scale for pairing. In our
system, Eq.~(\ref{Rheapprox}) can be written as

\begin{equation}
  R_{he}(L) \approx  \tanh^2[L \Gamma/(2 d_j)] \; , 
  \label{RheL}
\end{equation} 
with the saturation length $d_j/\Gamma$ as the length scale for pairing 
in the normal slab due to the proximity to the attached superconductor.

\section{Channel opening} 
\label{secopening}

Using Eqs.~(\ref{Gls}) and (\ref{arctan}) the NS linear conductance
can be calculated as function of $E_F$, $\Gamma$, $L$, and $W$.  We now
concentrate on a special case of these parameters, which results in
sharp resonances of the NS linear conductance as function of the Fermi
energy.  While in all the other cases considered in this paper we are
interested in the emergence of constructive interference that leads to
enhanced Andreev reflection and therefore consider cases in which
electrons in at least some of the channels hit the interface 
$N \gg 1/\Gamma$ times, we are now interested in a different limit, in 
which $\Gamma L \sqrt{k_F/W} \ll 1$.  The generic behavior in such a
structure would be that the current flow to the superconductor is
small, since the number of times an electron in any transverse channel
hits the barrier is smaller than $1/\Gamma$.  However, if we change
the Fermi energy (e.g. by a back gate) such that the channel of the
highest transverse mode has just opened, then the trajectory of a
particle in this channel is almost perpendicular to the interface. As
a result, the number of times a particle in this channel hits the
interface is much larger than $1/\Gamma$ and the contribution of this
channel to the NS conductance is significant and given by 

\begin{equation}
G_{31}^{sing}  = \frac{4 e^2}{h} 
\Theta(k_{j\|}^2) \tanh^2{\frac{j\pi L \Gamma}{4 W^2 k_{j\|}}} \; .
\label{Gsing}
\end{equation} 

\begin{figure} 
	\begin{center}  
	\begin{picture}(1,0.618)
	\put(-0.0,-0.0){\psfig{figure=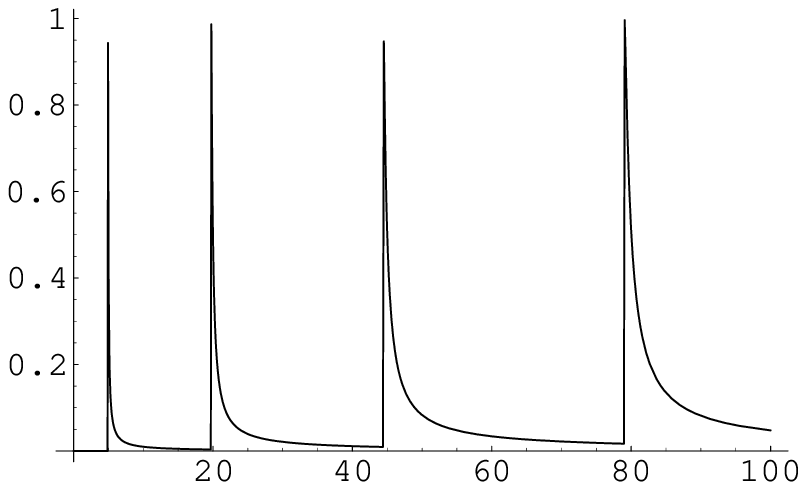,width=3in}}
	\put(0.0,-0.9){\makebox(0,0)[lb]{\psfig{figure=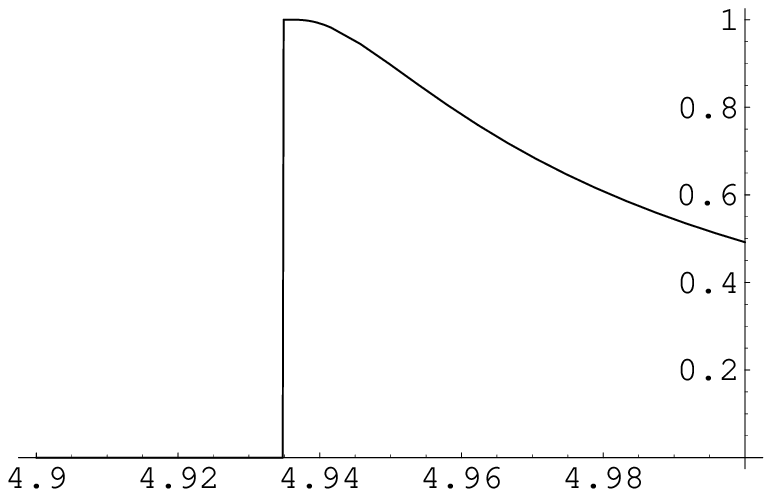,width=3in}}} 
	\put(0.1, 0.7){\makebox(0,0)[r]{{$G_{31}$}}}
	\put(0.9, -0.0){\makebox(0,0)[t]{{$E_F$}}}
	\put(0.95, -0.18){\makebox(0,0)[l]{{$G_{31}$}}}
	\put(0.95, -0.9){\makebox(0,0)[t]{{$E_F$}}}
	\put(0.5, -0.07){\makebox(0,0)[t]{{$(a)$}}}
	\put(0.5, -0.97){\makebox(0,0)[t]{{$(b)$}}}
	\end{picture} 
	\end{center}
	\vspace{8.0 cm}
	\caption{$(a)$ The conductance (in units of $4 e^2/h$) 
	between the left normal reservoir 
	and the superconductor is plotted as function of $E_F$ [in units of 
	$\hbar^2/(m W^2)$] for $L \Gamma/(4 W)=0.1$.  
	$(b)$ Enlarging the first peak we see that the conductance at the 
	peak is unity, and the width of the peak is approximately 
	$[L \Gamma/(4 W)]^2=0.01$ of the 
	value of $E_F$ at the peak.  } 
	\label{figchannelopen}
\end{figure}

\noindent 
In this equation we assumed that $\Gamma \ll 1$ and neglected the fact
that a fraction of the trajectories hit the barrier $N_j+1$ and not
$N_j$ times, since as the channel opens $N_j \gg 1$.  Defining
$\epsilon_j = \hbar^2 (k_{j\|})^2/(2 m)$ and 
$\tilde{\epsilon} = [L \Gamma/(4 W)]^2 E_F$ we find that for 
$0 < \epsilon_j \ll \tilde{\epsilon}$

\begin{equation}
G_{31}^{sing}  = \frac{4 e^2}{h} 
\tanh^2\sqrt{\frac{\tilde{\epsilon}}{\epsilon_j}} \approx   \frac{4 e^2}{h} 
[1-4e^{-2\sqrt{\tilde{\epsilon}/\epsilon_j}}]\; .
\label{Gsingapprox}
\end{equation} 
While in a normal quantum point contact connected to a superconductor
in series the linear NS conductance as function of the Fermi energy in
the slab would show steps \cite{Bee92}, in our case, where a
superconductor is attached to the point contact on its side, with a
barrier at the interface, the NS conductance as a function of $E_F$
has sharp {\it peaks} at the energies where channels open. The magnitude of
these peaks is $4$ times the quantum conductance, and the scale of their 
energy width is $\tilde{\epsilon}$. 
With the condition given above, $\Gamma L \sqrt{k_F/W} \ll 1$,  
these peaks are narrower than the energy difference 
between the opening of adjacent channels. For $L \approx W$ this condition 
reduces to  $\Gamma \ll \sqrt{\lambda_F/W}$. 
Under the semiclassical approximation we make 
the conductance peaks are
nonanalytical as a function of Fermi energy at energies where
transverse channels are opened, as can be seen both in
Eq.~(\ref{Gsingapprox}) and Fig.~\ref{figchannelopen}. This is due to
the fact that the number of times an electron hits the interface
diverges as $\epsilon_j \rightarrow 0$. 
These nonanalyticities are a
consequence of the semiclassical model, and one has to take into account 
that the validity of 
the semiclassical approximation is limited by the condition 
$k_{j\|} \gg 1/L^*$, 
where $L^*$ is the range of the potential variation. 
We estimate $L^*$ by $k_\perp/\nabla k_{\perp}=W*s/(\lambda_F \sqrt{\Gamma})$, 
taking into account the 
variation of $k_{\perp}$ with the spatial variation of $\Gamma$. 
The condition $k_{j\|} \gg 1/L^*$ is equivalent to 
$\epsilon_j \gg [\Gamma \lambda_F^4/(W^2 s^2)] E_F$, which is consistent 
with the condition $\epsilon_j \ll \tilde{\epsilon}$ given that 
$\Gamma \gg \lambda_F^4/(L^2 s^2)$. 
One can therefore expect that with this condition fulfilled, 
going beyond the 
semiclassical approximation would smoothen the above nonanalyticities, 
but will not alter the other features of the peaks (height and width). 
For $\epsilon_j \gg
\tilde{\epsilon}$ the contribution of the $j$th channel to the NS
conductance is proportional to $1/\epsilon_j$ and is given by
\begin{equation}
G_{31}^j = \frac{4 e^2}{h} \frac{\tilde{\epsilon}}{\epsilon_j} \; .
\label{Gjepsilon}
\end{equation} 

In this section we described the effect of channel opening on the NS
conductance of the system. The behavior of $G_T$ is similar, only the
peaks at the energies where channels open are half the magnitude and
are on top of the step function of magnitude $2e^2/h$ (since $G_T$ is
similar to $G_{31}$, only $2R^j_{he}$ is replaced by
$1+R^j_{he}$). The NN conductance is given by the complementary of
half the NS conductance to a step function ($2R^j_{he}$ is replaced by
$1-R^j_{he}$).  In the next sections we mostly consider $G_{31}$, and 
analogies to $G_T$ and $G_{21}$ can be made in a similar way.

\section{Magnetoconductance of a long normal slab attached to a superconductor}
\label{secmagnetic}

In this section we consider the same VNS structure as in
Secs.~\ref{secNtrajectory} and \ref{secopening} with a magnetic field
applied perpendicular to the slab.  We investigate the effect of the
magnetic field on the transmission probability of an $N$ trajectory,
as well as on the linear conductances in the system.  Magnetic field
penetration into the superconductor is neglected.

We consider $H \ll \Phi_0/(\lambda_F W)$ where $\Phi_0$ is the flux
quantum. Under this condition the curving of the trajectories of the
particles in the normal slab can be neglected \cite{Sch00}.  
The number of times, $N_j$ (or $N_j+1$), a trajectory of an electron in
channel $j$ hits the NS interface stays unchanged, and so do
Eqs.~(\ref{Glr})--(\ref{GT}), except that the Andreev reflection
probabilities $R_{he}(N)$ now depend also on $\Phi_H$, the phase
acquired by an electron and a hole moving in opposite directions
between two consecutive points the trajectory hits the NS interface
(``trajectory section'').  This phase is given by $\Phi_H=4 \pi H
A/\Phi_0$, where $A$ is the area of the triangle enclosed by the
trajectory section and the interface.

\begin{figure} 
	\begin{center}  
	\begin{picture}(1.1,0.68)
	\put(-0.02,-0.1){\psfig{figure=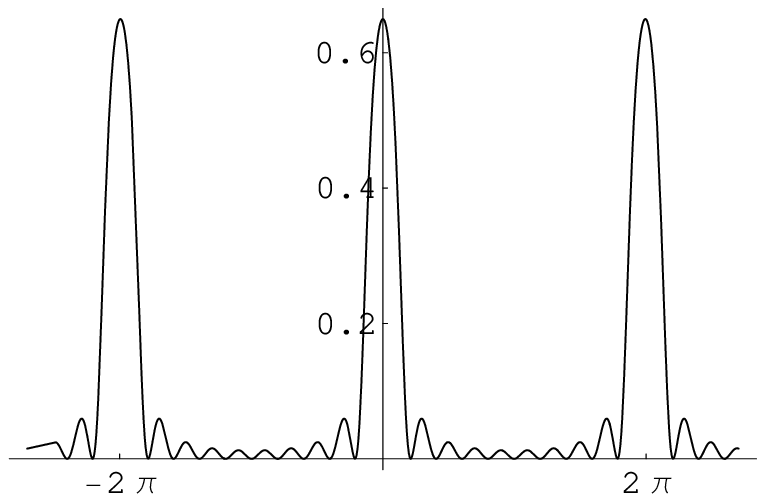,width=3.375in}}
	\put(-0.02,-1.1){\psfig{figure=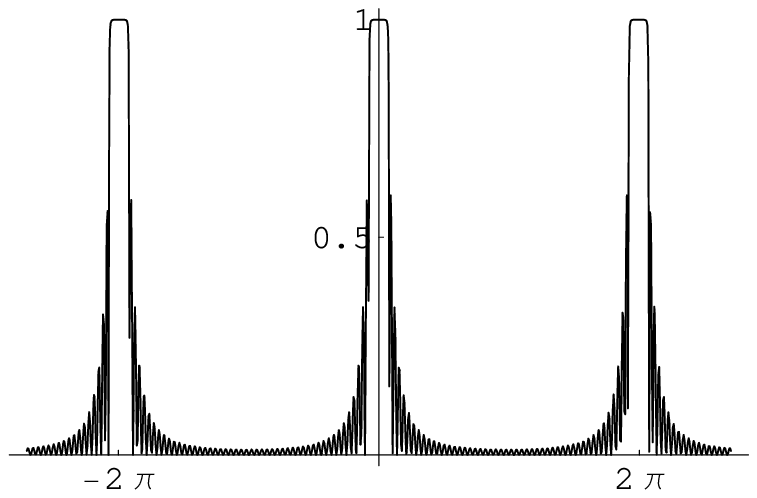,width=3.375in}}
	\put(0.54, 0.7){\makebox(0,0)[b]{{$R_{he}$}}}
	\put(1.08, -0.06){\makebox(0,0)[t]{{$\Phi_H$}}}
	\put(0.54, -0.3){\makebox(0,0)[b]{{$R_{he}$}}}
	\put(1.08, -1.06){\makebox(0,0)[t]{{$\Phi_H$}}}
	\put(0.54,-0.11){\makebox(0,0)[t]{{(a)}}}
	\put(0.54,-1.11){\makebox(0,0)[t]{{(b)}}}
	\end{picture} 
	\end{center}
	\vspace{8.7 cm}
	\caption{$R_{he}(N,\Phi_H)$ as obtained from 
	Eq.~(\protect\ref{RheNPhi}) for (a) $N=10$ trajectory and 
	(b) $N=50$ trajectory 
	in a VNS system (Fig.~\protect\ref{newvns}) 
	for a barrier transmission probability $\Gamma=0.2$. 
	Notice the narrow large peaks periodic in $\Phi_H$, 
	and the small oscillations between each such peaks, having 
	in (a) $N-1=9$ 
	nodes. In (b) the magnitude at the high peaks is approximately 
	unity, and the oscillations between them are hardly visible.
}
	\label{figRhemag1050_02}
\end{figure}

We now turn to the calculation of $R_{he}(N,\Phi_H)$. 
Repeating the same procedure leading to Eq.~(\ref{rheN}), but keeping 
track of the phase introduced by the magnetic field, we obtain 

\begin{eqnarray} 
r_{he}(N,\Phi_H) = 
r_{he} + \frac{r_{ee} r_{he}(N-1,\Phi_H) e^{i\Phi_H} 
r_{hh}}{1- r_{eh} r_{he}(N-1,\Phi_H) e^{i\Phi_H}} \; \; .
\label{rhephi}  
\end{eqnarray} 
In order to obtain an explicit formula for $r_{he}(N,\Phi_H)$ it is 
useful to define 

\begin{equation}
r_{he}(N,\Phi_H)=r_{he} \frac{\beta_N}{\gamma_N} \; , 
\label{betagamma}
\end{equation} 
where $\beta_N$ and $\gamma_N$ are also $\Phi_H$ dependent. 
Inserting this definition for $N$ and $N-1$ into Eq.~(\ref{rhephi}) we 
obtain the matrix equation 

\begin{eqnarray}
\left( \begin{array}{c}\beta_N \\ \gamma_N \end{array} \right) & = & 
\left( \begin{array}{cc}e^{i \Phi_H} & 1  \\ r^2 e^{i \Phi_H} & 1
\end{array} \right) 
\left( \begin{array}{c}\beta_{N-1} \\ \gamma_{N-1} \end{array} \right) 
\nonumber \\ & = & 
\left( \begin{array}{cc}e^{i \Phi_H} & 1  \\ r^2 e^{i \Phi_H} & 1
\end{array} \right)^{N-1} 
\left( \begin{array}{c} 1 \\ 1 \end{array} \right) \; . 
\label{matrixmagnetic} 
\end{eqnarray}  
Here we used the fact that at the Fermi energy $r_{he}$ is imaginary.
$\beta_N$ and $\gamma_N$ are in principal defined up to (the same) 
multiplication constant, which we dictate by the choice $\beta_1=\gamma_1=1$.
Diagonalizing the matrix and taking it to the power $N-1$ we obtain 

\begin{eqnarray} 
\label{rheNPhi} 
&&r_{he}(N,\Phi_H) =  \\ 
&&\frac{i r e^{-i\Phi_H/2}}{-i\sin{(\Phi_H/2)} + 
\sqrt{b} \coth{\left[N \tanh^{-1}{\left(\displaystyle\frac{\sqrt{b}}
{\cos{(\Phi_H/2)}}\right)}\right]}} \; , \nonumber
\end{eqnarray} 
where $b=r^2 - \sin^2{(\Phi_H/2)}$. 
Using the fact that the second term in the denominator is always 
real (also when $b$ is negative) we obtain 

\begin{eqnarray}
\label{RheNPhi}
&&R_{he}(N,\Phi_H) =  \\ 
&&\frac{r^2}{\sin^2{(\Phi_H/2)} + 
b \coth^2{\left[N \tanh^{-1}{\left(\displaystyle\frac{\sqrt{b}}
{\cos{(\Phi_H/2)}}\right)}\right]}} \; , \nonumber 
\end{eqnarray} 
which, for $b<0$, can be written as 


\begin{eqnarray} 
\label{RheNPhinegb}
&&R_{he}(N,\Phi_H) = \\ 
&&\frac{r^2}{\sin^2{(\Phi_H/2)} + 
(-b) \cot^2{\left[N \tan^{-1}{\left(\displaystyle\frac{\sqrt{-b}}
{\cos{(\Phi_H/2)}}\right)}\right]}} \; . \nonumber 
\end{eqnarray} 
These formulas hold for any $N$, $r (\Gamma)$, and $H$. 
We now consider the case of $r \ll 1$ and $N r \gg 1$. 
For $\sin^2{(\Phi_H/2)} \gg r^2$ 
(which holds for most values of $\Phi_H$ when $r \ll 1$), 
we see from Eq.~(\ref{RheNPhinegb}) that 
$R_{he}(N,\Phi_H) \ll 1$. In the opposite limit, of 
$\sin^2{(\Phi_H/2)} \ll r^2$, one obtains from Eq.~(\ref{RheNPhi}) that 
$R_{he}(N,\Phi_H) \approx 1$. This leads to the conclusion that 
the Andreev reflection probability from an $N$ trajectory is small 
for almost all values of perpendicular magnetic field, except those special 
values that result in $|\Phi_H - 2 k \pi| \lesssim r$ 
(Fig.~\ref{figRhemag1050_02}). 
Between every two such peaks the function 
oscillates, having $N-2$ smaller peaks (and $N-1$ nodes). These peaks 
(nodes) correspond to integer (half integer) flux quanta through an area 
of an integer number of triangles in the trajectory of the specific channel. 

The magnetic field not only impairs the constructive interference 
leading to large Andreev reflection at zero field, but causes 
destructive interference.  
This can be seen by considering $\Phi_H = \Phi_0/2$. 
Then, $R_{he}(N)=0$ for even $N$ and for
odd $N$ it equals $r^2$, the Andreev reflection probability from a
single hit. Therefore, for any given channel $R_{he}^j$ is of order
$r^2$.  
In the cases discussed in Sec.~\ref{pairinglength}, of no interference
and of constructive interference, there were (different) saturation lengths
beyond which the Andreev reflection probability was close to unity. 
Here, however, due 
to destructive interference, there is no such length scale and Andreev 
reflection is small (~$r^2$) {\it for any length of NS interface.} 
This is true also when the destructive interference is due to a
superconducting phase difference in an SNS structure, as is discussed
in Sec.~\ref{secSNS}. 

\begin{figure} 
	\begin{center}  
	\begin{picture}(1,0.618)
	\put(-0.05,-0.2){\psfig{figure=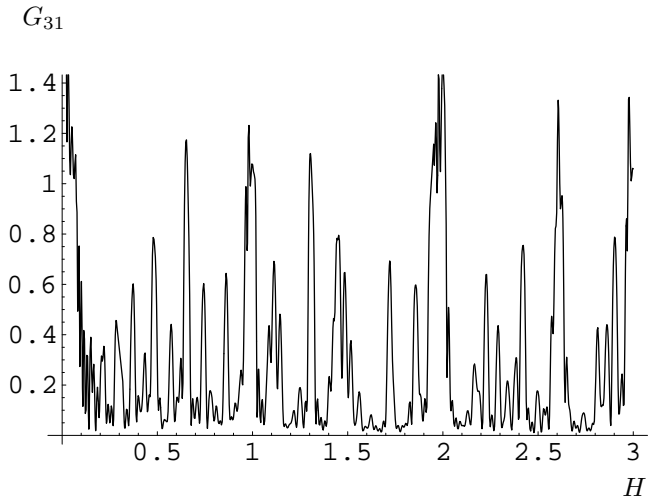,width=3.375in}}
	\put(0.02, 0.57){\makebox(0,0)[b]{{$G_{31}$}}}
	\put(1.05, -0.22){\makebox(0,0)[t]{{$H$}}}
	\end{picture} 
	\end{center}
	\vspace{2.0 cm}
	\caption{The conductance (in units of $4 e^2/h$) 
	between the left normal reservoir 
	and the superconductor as function of $H$ 
	(in units of $\Phi_0/W^2$). 
	The plot is given for $L/W=50$, $\Gamma=0.1$ and 
	$2 m E_F W^2/\hbar^2 = 1000$, which corresponds to 10 open channels.
	Peaks higher than unity are a result of overlapping resonances 
	of 2 or more channels. Apparent periods of $H$ are: 
	0.29, 0.37, 
	0.48, 0.65. At $H = 0$ the peak is significantly higher than 
	the others ($G_{31}=5$, not shown).}
	\label{figmagneticpeak1}
\end{figure}

Inserting Eq.~(\ref{RheNPhi}) in the conductance
formulas [Eqs.~(\ref{Glr})-(\ref{GT})] we obtain the linear conductances as
function of magnetic field for any junction parameters
($\Gamma,L,W,E_F$).  If the parameters of the structure are such that
$\Gamma L/W \gg 1$, then at zero magnetic field the NS dimensionless
conductance is much larger than unity, since there are many channels
for which $N \Gamma \gg 1$.  We choose such a case and plot in
Fig.~\ref{figmagneticpeak1} the NS dimensionless conductance as a 
function of $H$.  At $H=0$ the conductance has a sharp peak, of mag-

\noindent 
nitude of the order of the number of channels and width of order
$r$.  As $H$ is increased, the constructive interference leading to
the enhanced Andreev reflection is destroyed in one channel after the
other and the NS conductance becomes small. However, the conductance
of each channel is periodic in $H$, with a period given by the area of
the triangle between a trajectory section in this channel and the
interface. This quasiperiodicity is reflected in the peak spectrum of
the NS conductance as function of $H$ (Fig.~\ref{figmagneticpeak1}).
Periods of larger $H$ reflect channels with a smaller triangle, which
corresponds to a trajectory of larger $N$, and therefore [see
Eq.~(\ref{Rheapprox})] the peak heights are larger.  Using this
quasiperiodicity one can obtain ``magnetic switching.'' By choosing
the mag-

\noindent
netic field such that $|\Phi_H^j = 2k\pi|$ for one channel
only, one can remove only electrons propagating in this channel from
the normal slab to the superconductor and have the electrons in all
the other channels propagate to the right reservoir with probability
close to unity.

Though the results in this section were given for the linear
conductance at finite magnetic field, it is straightforward to
generalize our results to be valid for finite subgap voltage, thus
obtaining the differential conductance as a function of voltage. One can
also incorporate a constant phase gradient $\nabla \phi$ in the
superconductor in parallel to the NS interface, generated by a
constant supercurrent.  The Andreev reflection amplitude from an $N$
trajectory would then be given by Eq.~(\ref{rheNPhi}) with $\Phi_H$
replaced by

\begin{equation}
\Phi_{tr} = \Phi_H + \nabla \phi \, d_j + \Phi_j^{tr}(\epsilon) \; . 
\label{Phitr}
\end{equation} 
Here $\Phi_j^{tr}(\epsilon) = (k_{j\|}^+ -
k_{j\|}^-) W/\sin\theta_j + \arg[r_{ee}r_{hh}]$ is the relative
electron-hole phase accumulated due to finite energy in one triangle,
where $k_{j\|}^\pm = k_{j\|}(E_F \rightarrow E_F \pm \epsilon$).  Note
that there is a complete analogy between applying a perpendicular magnetic
field and constant gradient of the superconducting phase, with the
relation $\nabla \Phi = 2 \pi H W / \Phi_0$.  For $H=0$ and $\nabla
\Phi = 0$, we obtain a zero-bias peak in the differential NS conductance
as a function of voltage, similar to the low-$H$ behavior of the NS
conductance as shown in Fig.~\ref{figmagneticpeak1}.

\section{Conductance parallel to the interface in an SNS system}
\label{secSNS}

We now consider a system in which a second superconductor is attached
symmetrically to the other side of the slab, so the structure has
up-down symmetry.  We consider the case in which the two barriers
between the normal slab and the superconductors have the same
transmission probability $\Gamma$.  Then, one can apply the same
approach we used in the previous sections, only count the number of
hits of each trajectory at both interfaces.  The calculation of
$R_{he}(N,\Phi_s)$ is given in Appendix~\ref{appSNS}.  It is done in the
same spirit as the calculation of $R_{he}(N,\Phi_H)$, but is more
elaborate since one has to distinguish between even and odd times a
trajectory hits the interface, and the recursion relations are more
complicated.  The result for even $N$ is 

\end{multicols}
\widetext

\noindent
\begin{picture}(3.375,0)
  \put(0,0){\line(1,0){3.375}}
  \put(3.375,0){\line(0,1){0.08}}
\end{picture}
\begin{equation}
R_{he}(2N,\Phi_s)  =   
\left(z + (1-z) \coth^2{\left[N 
\tanh^{-1}\left({\frac{2r \cos(\Phi_s/2) \sqrt{(1-z)}}
{1+r^2\cos{\Phi_s}}}\right)\right]} 
\right)^{-1} 
\label{evenSNS} 
\end{equation} 
\hfill
\begin{picture}(3.375,0)
  \put(0,0){\line(1,0){3.375}}
  \put(0,0){\line(0,-1){0.08}}
\end{picture}

\begin{multicols}{2}
\noindent
where $z=r^2\sin^2{(\Phi_s/2)}$. 
The result for odd $N$ is similar and is given in Appendix~\ref{appSNS}. 
These results are even in $\Phi_s$ and, therefore, the same for a trajectory 
hitting first either of the two superconductors. 
For $r \ll 1$ 
we obtain $R_{he}(2N,\Phi_s) = \tanh^2{\left[2 N r \cos(\Phi_s/2)\right]}$. 
For $N r \gg 1$ we see that $R_{he}(2N,\Phi_s) \approx 1$ unless
$|\Phi_s - (2k+1)\pi| \lesssim \pi/(N r)$, while $R_{he}(2N,\Phi_s) =
0$ for $\Phi_s = (2k+1)\pi$. As a result, in this limit, the Andreev
reflection probability from a $2N$ trajectory as function of $\Phi_s$
has sharp {\it dips} near $\Phi_s = (2k+1)\pi$ of an approximate width of 
$1/(2 N r)$ (Fig.~\ref{figSNSeven}). The transmission probability of
electrons to the right reservoir is given by $1 - R_{he}(2N,\Phi_s)$
and has sharp resonant peaks, which indicates that there is a
transverse Andreev level shifted to $E_F$ at $\Phi_s = (2k+1)\pi$,
similar to the case in standard SNS junctions \cite{WS96}.  As in the
case of perpendicular magnetic field, $R_{he}(2N,\Phi_s)$ has a
maximum at $\Phi_s=0$ (or multiples of $2\pi$), but there are two
major differences between the dependence of $R_{he}(N)$ on $H$ and on
$\Phi_s$: (i) In a period of $2\pi \; R_{he}(N,\Phi_s)$ has one
minima, where $R_{he}(N,\Phi_H)$ has $N$ minima. (ii) while
$R_{he}(N,\Phi_H)$ exhibits sharp {\it peaks} near $\Phi_H=2\pi n$,
and for most values of magnetic field (generically) constructive
interference is lost, and Andreev reflection is small, the situation
for $R_{he}(N,\Phi_s)$ is opposite. It is close to unity for most
values of phase difference, and exhibits sharp {\it dips} near
$\Phi_s=\pi+2\pi k$.


\setlength{\unitlength}{3in}
\begin{figure} 
 	\narrowtext 
	\begin{center}  
	\begin{picture}(1,0.618)
	\put(-0.05,-0.2){\psfig{figure=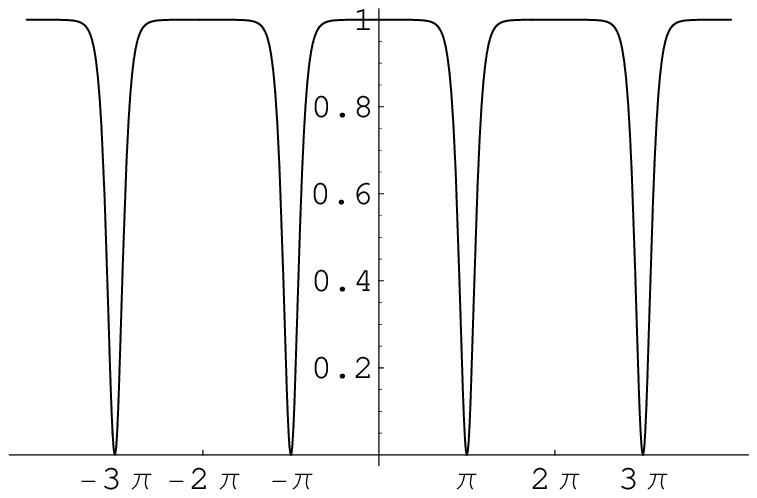,width=3.375in}}
	\put(0.51, 0.6){\makebox(0,0)[b]{{$R_{he}$}}}
	\put(1.03, -0.16){\makebox(0,0)[t]{{$\Phi_s$}}}
	\end{picture} 
	\end{center}
	\vspace{1.5 cm}
	\caption{$R_{he}(N,\Phi_s)$ in an SNS structure for $N=50$  
	and $\Gamma=0.2$. Note the narrow dips, in comparison to the 
	narrow peaks in Fig.~\protect\ref{figRhemag1050_02}(b), which 
	is drawn for the same $N$ and $\Gamma$, as a function of $H$.}
	\label{figSNSeven}
\end{figure}

These differences can be understood by examining the two mechanisms of
the destruction of the constructive interference between different
paths of the same trajectory that result in a hole returning to the
left reservoir. Indeed, the phase difference between a hole resulting
from an Andreev reflection at the first hit of the NS interface and a
hole resulting from an Andreev reflection at the second hit of the NS 
interface is similar in both cases ($\Phi_H$ and $\Phi_s$), but the
phase difference between this hole and a hole resulting from an
Andreev reflection at the $N$th hit of the NS interface is very
different for the two cases.  It is $(N-1) \Phi_H$ for the case of a
magnetic field and $\Phi_s$ or $0$ (depending if $N$ is even or odd)
for the case of superconducting phase difference.  This introduces a
large amplification factor in the electron-hole phase difference
introduced by the magnetic field compared to that introduced by the
superconducting phase difference.  As a result, the magnetic field is
far more efficient in destroying the constructive interference leading
to the enhanced Andreev reflection. 

\begin{figure} 
	\begin{center}  
	\begin{picture}(1,0.618)
	\put(-0.05,-0.2){\psfig{figure=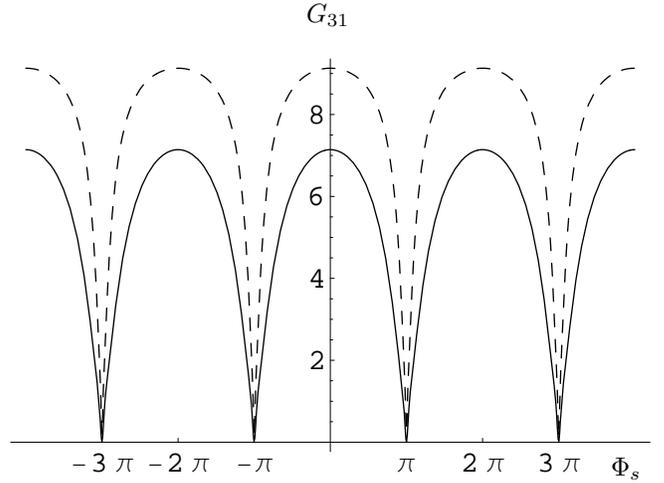,width=3.375in}}
	\put(0.51, 0.6){\makebox(0,0)[b]{{$G_{31}$}}}
	\put(1.03, -0.16){\makebox(0,0)[t]{{$\Phi_s$}}}
	\end{picture} 
	\end{center}
	\vspace{1.1 cm}
	\caption{$G_{31}$ (in units of $4 e^2/h$) as function of $\Phi_s$. 
	The solid curve is given for the same parameters as in 
	Fig.~\protect\ref{figmagneticpeak1} 
	to enable comparison. In contrast with the case of applied magnetic 
	field, the conductance is periodic in $\Phi_s$, has narrow dips at 
	odd multiples of $\pi$, and the oscillations as function of $\Phi_s$ 
	are of the order of the full conductance (giant).
	The dashed line is plotted for the same parameters,  
	except here $\Gamma=0.25$. The larger conductance and the 
	narrower dips are both a result of $N \Gamma$ being larger for each 
	channel.}
	\label{figSNSconductance1}
\end{figure}

The linear conductances as function of $\Phi_s$ are calculated  
by inserting the results for the Andreev reflection 
probabilities [Eqs.~(\ref{evenSNS}) and (\ref{oddSNS})] 
in the conductance formulas[Eqs.~(\ref{Glr})--(\ref{GT})]. 
$N_j$ in the VNS structure is now replaced by 
$\hat{N_j}$ which equals the integer part of $L \tan{\theta_j}/W$ and 
$p_j$ is replaced by $\hat{p_j} = L \tan{\theta_j}/W - \hat{N_j}$. 
In Fig.~\ref{figSNSconductance1} we plot 
(solid line) the NS conductance as a function 
of the superconducting phase difference for a system with the same parameters 
as the one in Fig.~\ref{figmagneticpeak1}, for comparison. 
The NS conductance for a similar 
structure with a larger barrier transmission probability ($\Gamma=0.25$) 
is also plotted (dashed line) to demonstrate the narrowing of the width of 
the dips as $N \Gamma$ grows. 
Here we see another marked 
difference between applying a perpendicular magnetic 
field in the VNS structure and a phase difference in the SNS structure. 
In the case of the applied magnetic field 
there is a large peak at $H=0$, to which 

\noindent 
all the channels contribute 
due to the constructive interference, but the 
periodic peaks at higher fields appear for each channel at a different $H$. 
If the parameters of the junction are such that the separations between 
conductance peaks are smaller than their width, the result will be a smooth 
oscillatory behavior of the conductance as function of $H$.
On the other hand, in the case of SNS structure, the dips at all 
$\Phi_s = (2k+1)\pi$ are common to 
all the channels, and therefore the conductance oscillations as function 
of $\Phi_s$ show sharp features of magnitude of the 
order of the total conductance. 

Recently, Petrashov {\it et al.} measured large conductance
oscillations as a function of the magnetic field and superconducting phase
difference in a normal slab connected to ``superconducting mirrors''
\cite{PADC93,PADC95}.  Our results cannot be directly applied to the
experimental structures studied by Petrashov {\it et al.}, since in the
experiment the structures are different, the superconductor is floating, 
and there is finite scattering in the normal slab. 
However, some features appear
to be more general and exist both in the experimental results and in our
calculations. These are the much larger magnitude and sharpness 
of the oscillations
as a function of superconducting phase difference compared to the
oscillations as a function of magnetic field. 

\begin{figure} 
	\centerline{\psfig{figure=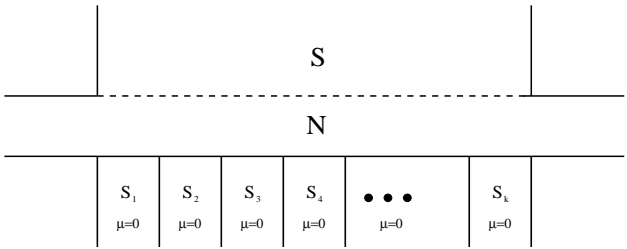,width=3.375in}}
	\vspace{0.3cm}
	\caption{SNS junction where the bottom superconductor is divided to 
	pieces with controllable phase of the order parameter (only the 
	relevant region is shown). }
	\label{figmultiples}
\end{figure}

To conclude this section we now apply the results obtained above to show how 
one can get controlled current withdrawal from an electronic waveguide. 
Here we use the fact  that as long as the phase difference between the 
superconductors is 
$\pi$, the electrons move in the slab as in a waveguide.
By replacing the bottom superconductor with a series of superconductors 
(Fig. \ref{figmultiples}), 
each with a controllable phase $\phi_i$ and interface length with the slab, 
$\tilde{L}$, we can create a ``switch'' in 
which we can control the location where the current is drawn. We set 
$\phi_{i \neq \tilde{i}} = \pi$ and $\phi_{\tilde{i}} = 0$. 
For all $i \neq \tilde{i}$ an incoming electron from the left in an angle with 
$\tan{\theta} \gg (2W/\tilde{L} \Gamma)$ 
will be normally reflected as in a waveguide. 
However, when it reaches the $\tilde{i}$th superconductor Andreev reflection 
occurs, 
adding a Cooper pair to the superconductor. 
One can therefore inject a current from the left 
reservoir and draw it at any one of the superconducting slabs. 

\section{Shot noise}
\label{secnoise} 

In this section we calculate the shot noise as function of $H$ 
in the VNS structure and $\Phi_s$ in the SNS structure, and show that 
the differences between applying magnetic field and superconducting 
phase difference are reflected remarkably in the shot noise properties 
of the systems.  

We define the quantities $P_{ll'} = 2 \int_{- \infty}^{\infty} dt  
\langle \Delta \hat{I}_l(t) \Delta \hat{I}_{l'}(0) \rangle$ 
(where $l,l'$ are normal terminal indices), which give both the shot
noise and the cross correlators between current fluctuations at the
two normal terminals.  Anantram and Datta \cite{AD96} considered the
case where an arbitrary number of superconducting and normal terminals
exist, with the restriction that the chemical potential in all the
superconducting terminals is the same, and obtained general equations
for the current correlators.  Using their equations for our system we
obtain 

\begin{equation}
P_{11} = P_{22} = P_{12} = P_0 \sum_j \Theta(k_{j\|}^2) 
R^j_{he}(1-R^j_{he})  \; , 
\label{noise1ch}
\end{equation}
where $P_0=2e|V|(2e^2/h)$. This formula is applicable for both the VNS and 
SNS structures, and the specific parameters of the junction as well as the 
$H$ and $\Phi_s$ dependence enter only into $R^j_{he}$. Notice the full 
positive noise correlations between the two normal terminals \cite{TML01}, 
which is a 
result of zero normal reflection to the same reservoir and Andreev 
transmission to the other reservoir in our model. 

Due to the dependence of the shot noise on the functions
$R^j_{he}(1-R^j_{he})$, it shows peaks at points where the Andreev
reflection amplitude is neither close to zero or unity.  For the SNS
structure we consider, this results in

\begin{figure} 
	\begin{center}  
	\begin{picture}(1,0.618)
	\put(-0.05,-0.2){\psfig{figure=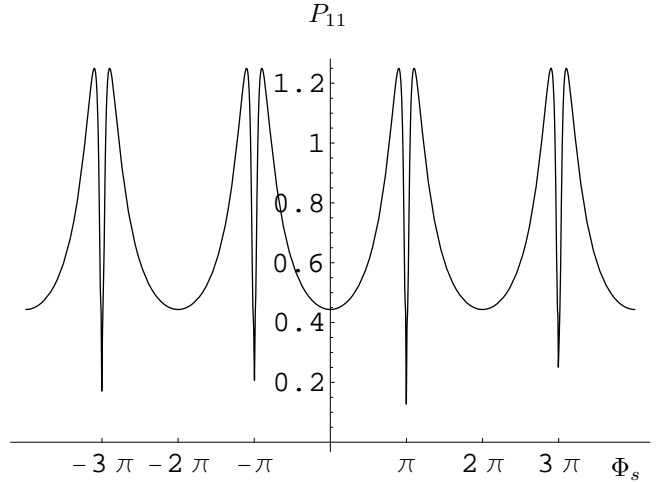,width=3.375in}}
	\put(0.51, 0.6){\makebox(0,0)[b]{{$P_{11}$}}}
	\put(1.03, -0.16){\makebox(0,0)[t]{{$\Phi_s$}}}
	\end{picture} 
	\end{center}
	\vspace{1.1 cm}
	\caption{$P_{11}$ in units of $P_0$ as function of $\Phi_s$ 
	for the geometry where the normal slab is attached to 
	two superconductors. The parameters of the system are the same 
	as those in Fig.~\protect\ref{figSNSconductance1} (dashed line). }
	\label{figSNSnoise25}
\end{figure}

\noindent 
a sharp feature near the values
of $\Phi_s$ corresponding to {\it dips} in the NS conductance [$\Phi_s
= (2k+1)\pi$], which are common to all the channels.  The form of the
sharp 
feature is two double peaks separated by a very sharp dip
\cite{FLB98}, as can be seen in Fig.~\ref{figSNSnoise25}.  As function
of $H$, in the case of one channel, a sharp feature appears near each
value of $H$ corresponding to a {\it peak} in the conductance.
However, these points are channel dependent, and as was the case for
the conductance, the presence of many channels smears the sharp
features as function of $H$. 

Our results for the noise are easily generalized to finite subgap
energy (differential shot noise as function of bias voltage) and
gradient of the superconductor phase in the same manner discussed at the 
end of Sec.~\ref{secmagnetic}.

\section{Diffusive NS junctions} 
\label{secdiffusive}

The semiclassical approach, introduced by van Wees {\it et al.} to explain 
the phenomenon of reflectionless tunneling in diffusive NS junctions 
\cite{WVMK92} was used to analyze numerically other experimental results 
as well (see, e.g., Refs.~\cite{MHW+97} and \cite{MWK+97}). 
This approach has proved useful in obtaining a qualitative understanding 
of the physical phenomena in various geometries which make the use of 
the standard methods (quasiclassical formalism, BdG equations) difficult. 
As Eq.~(\ref{Rheapprox}) applies to any NS system, ballistic or diffusive, 
it can be used to obtain analytical results within this approach. 
In this section we apply the semiclassical formalism to 
treat both the phenomenon of reflectionless tunneling and the reduction 
of the local density of states (DOS) across a diffusive NS junction. 
We show that both of these phenomena are a result of the large transparency 
of the barrier to pair current (although $\Gamma \ll 1$ and under the 
conditions given by the general criterion at the end of Sec.~\ref{NAndreev}) 
and thus stem from the same physical effect. 

\begin{figure} 
	\centerline{\psfig{figure=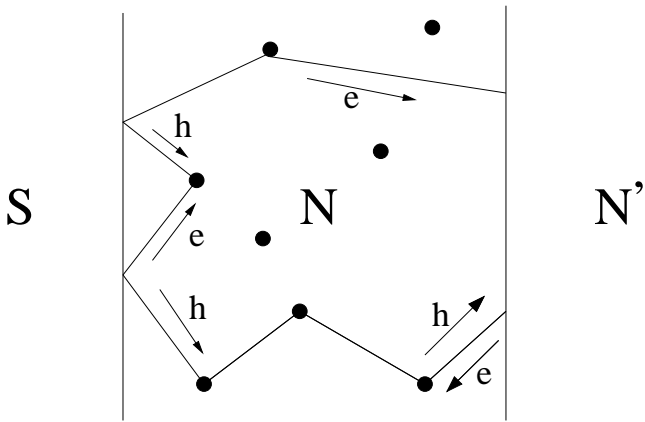,width=3.375in}}
	\caption{Geometry of the model \protect\cite{WVMK92}, 
	an example of a trajectory with $N$=2.}
	\label{fig-van}
\end{figure}

\noindent 
We also demonstrate the connection between the 
effect of reflectionless tunneling in ballistic systems discussed in this 
paper, to the one in diffusive systems. 

Since we consider the particle's trajectory in the normal metal to be
deterministic, our approach can be expected to give correct
quantitative results [Eqs.~(\ref{eq:Fullconductance}) and 
(\ref{eq:Glimits})] when the scattering potential varies slowly on a
scale of a wavelength \cite{WVMK92,BH91}, and the sample is short
enough such that classical dynamics will not develop phase space
structures on scales smaller than $\hbar$ \cite{Argaman}.  For short
range disorder, this approach is expected to give results which are
qualitatively correct \cite{WVMK92}.

Following the treatment of Ref.~\cite{WVMK92}, but using our
analytical result for the Andreev reflection probability from an $N$
trajectory, Eq.~(\ref{arctan}), we obtain the linear conductance of a
normal slab connected via a barrier to a superconducting reservoir
(see Fig.~\ref{fig-van}), which for $\Gamma \ll 1$ is given by

\begin{equation} 
	G(V \rightarrow 0,H=0)=\frac {4 e^2 n}{h} 
	\sum_{N=0}^{\infty} T^2 (1-T)^{N-1} \tanh^2(N \Gamma/2 ) . 
  \label{eq:Fullconductance} 
\end{equation}
Here $n$ is the number of channels and $T$ is the average transmission 
probability of the normal slab (the mean free path divided by its length).
In the two limits where $\Gamma \ll T$ and $\Gamma \gg T$ 
Eq.~(\ref{eq:Fullconductance}) reduces to 

\begin{equation}
  G(V \rightarrow 0,H = 0) = \left\{ \begin{array}{ll} 
      \frac {2 e^2 n}{h} \frac{\Gamma ^2}{T}  &  (\Gamma \ll T) \; , \\ 
	 \\ 
      \frac {2 e^2 n}{h} (\frac{1}{2T} + \frac{1}{\Gamma})^{-1}  &  (\Gamma 
      \gg T) \; . 
    \end{array}   
  \right. 
  \label{eq:Glimits}
\end{equation} 

Unlike ballistic systems, in this case multiple reflections from the 
interface are enabled by the disorder. 
For small disorder $\Gamma \ll T$, the conductance is proportional to
$\Gamma^2$, reflecting the fact that Andreev reflection is a 
two-particle process. Already in this limit the disorder increases the
conductance by a factor of $1/T$.  For $T \ll
\Gamma$ the disorder is large enough to generate, with high probability, 
trajectories with $N \gg 1/\Gamma$, as we show below, and therefore  
the barrier is not effective and the conductance is linear in $T$. 
The conductance has a maximum for $T
\approx \Gamma$, where $G \approx (2 e^2 n/h) \Gamma \approx (2 e^2
n/h) T$. 

Equation~(\ref{eq:Glimits}) 
differs in the $\Gamma \gg T, T \rightarrow 0$ limit by a factor of $2$ 
from the analogous formula 
obtained by Beenakker {\it et al.} \cite{BRM94} for short-range disorder.  
A detailed discussion of how this discrepancy results from the different 
assumptions in the two models is given in Appendix~\ref{appfactor2}. 

As the  essence of the effect of reflectionless tunneling is the fact 
that the barrier, though high, is transparent to pair current, the condition 
for the barrier to be ineffective was discussed for this phenomenon 
\cite{Vol94,ZSZ95} as well as for other phenomena, as the reduction of the 
DOS on the normal side of an N-insulator-S (NIS) semi-infinite junction 
\cite{Vol94}. We now show, using random walk theory, 
that the criterion stated in Sec.~\ref{NAndreev}, 
for the barrier to be ineffective, can be reduced in the diffusive case to 
the different known conditions for each phenomenon. 

In Appendix~\ref{app1} it is shown that the typical length of a diffusive
trajectory between $N$ consecutive times it hits the barrier is 
$L_N \approx N^2 l_n$, where $l_n$ is the elastic mean free path in the
normal metal (interestingly, there is no average length for an $N$
trajectory; see Appendix~\ref{app1}). This is also the order of magnitude
of the length of the longest loop in such a trajectory (a loop is a
part of a trajectory between two consecutive points it hits the
interface).  Using this result for $L_N$ and since large
contributions to Andreev reflection arise from trajectories that hit
the interface $N \gg \Gamma^{-1}$ times before losing electron-hole 
coherence (\ref{Rheapprox}),
only when coherent trajectories with total lengths larger than
$L_{\Gamma} \equiv l_n/\Gamma^2$ occur with high probability will the
barrier not be effective.  This requires the electrons and holes to be
coherent over a distance $\sqrt{L_{\Gamma} * l_n} = l_n/\Gamma$ from
the interface.  Therefore, the general condition in diffusive systems
for the barrier to be ineffective is $\xi \gg l_n/\Gamma$, where $\xi$
is the distance from the interface at which electrons and holes are
still coherent.  The coherence distance $\xi$ is determined by the
length of the slab, energy of the electron, or magnetic field,
depending on the physical case considered.
When measuring the conductance of an NS junction, then for 
zero energy and zero magnetic field, the length of the normal
metal, $d$, is what limits the trajectories to lengths of order
$d^2/l_n$  (since a particle that reaches a distance $d$ 
from the interface enters the reservoir, where phase coherence is lost). 
Therefore, the barrier is not effective when $\Gamma \gg l_n/d$.
Since the transmission probability through the diffusive normal part is 
roughly $T \approx l_n/d$, this condition reduces to the known condition  
\cite{Vol94} for the barrier to be ineffective, $\Gamma \gg T$ [in 
accordance with Eqs.~(\ref{eq:Fullconductance}) and (\ref{eq:Glimits})].

Following the same considerations one can obtain the condition for the
barrier to be ineffective in various cases, noting the different
mechanism impairing electron-hole coherence in each case.  We
consider, for example, the local DOS in a semi-infinite NS junction.
The reduction of the local DOS in the normal side of an NS interface
is closely related to the averaged amplitude of an electron near the
interface to return to the same point as a hole through the pair
amplitude $\langle \psi_\downarrow \psi_\uparrow \rangle$
\cite{Sch00}.  At zero energy and assuming the normal metal is 
semi-infinite, there is no mechanism that limits the length of coherent
trajectories. The electron hits the barrier as many times as needed,
$N \gg \Gamma^{-1}$, without losing electron-hole phase coherence, and
according to Eq.~(\ref{Rheapprox}), it is finally Andreev reflected.
This results in a finite pair amplitude throughout the normal part,
even in the presence of a barrier (at $\epsilon = 0$ there is no
reduction of the pair amplitude due to phase averaging), and in a zero
DOS at zero energy.

At finite energy $\epsilon$, the electron and hole moving in opposite
directions in a trajectory of length $L$ accumulate a relative phase
of $L \epsilon/(\hbar v_F)$ which limits the length of coherent
trajectories to order $\hbar v_F/\epsilon$, and to distance $\xi=\xi_n
\equiv \sqrt{\hbar D_n/\epsilon}$ from the interface (trajectories
that traverse a distance longer than $\xi_n$ result in phase
difference of order $2\pi$).  The condition for having a large Andreev
reflection amplitude is therefore $\Gamma \gg l_n/\xi_n$ \cite{Vol94}.
This condition assures that the total phase accumulated by the ingoing
electron and outgoing hole is less than $2 \pi$, and therefore the
averaging in $\langle \psi_\downarrow \psi_\uparrow \rangle$ results
in finite pair amplitude and the local DOS is reduced.  Similar
considerations result in the width of the zero-bias anomaly in
reflectionless tunneling being proportional to the Thouless energy 
\cite{Sch00}.

\section{Conclusions}

We have demonstrated the effect of reflectionless tunneling in a
ballistic NS system in which the multiple reflections from the
interface are due to the geometry.  We considered a normal slab with
superconductors attached to its sides, so the normal current flows in
parallel to the NS interface.  The barrier at the NS interface was
taken to be smooth, so that normal reflection is specular, and with
transmission probability $\Gamma \ll 1$.  We obtained a formula for
the Andreev reflection amplitude from a trajectory that hits the
barrier at the NS interface $N$ times and showed that, when 
$N \gg \Gamma^{-1}$, the barrier is transparent to pair current (though
$\Gamma \ll 1$), leading to good proximity.

We have shown that having a smooth rather than rough barrier at the
interface is advantageous in giving rise to more pronounced and
delicate features, which are not averaged over.  This results in new
measurable phenomena, such as the sharp {\it peaks} in the NS
conductance as new channels open (in contrast to the usual step
function) and quasiperiodicity of the conductance as function of
magnetic field $H$.  The smoothness of the barrier also enables one to
conduct detailed manipulations such as extracting out a single channel
from a normal metal (semiconductor) waveguide or extracting the
current at a given position along the waveguide. 

By obtaining explicit formulas for the three-terminal conductances of
the system as function of $H$ and of the superconducting phase
difference $\Phi_s$, we have shown that both $H$ and $\Phi_s$ impair
the constructive interference leading to the enhanced NS conductance,
but in a qualitative different way. While as a function of $H$ the
enhanced NS conductance is limited to a small range of magnetic field 
and is channel specific, as a function of $\Phi_s$ the enhanced NS
conductance is generic, and is destroyed only near specific values of
$\Phi_s$ (odd multiples of $\pi$) for all the channels, leading to
giant conductance oscillations.  This difference is also reflected
clearly in the shot noise behavior as function of both quantities.

By demonstrating the possibility to obtain large Andreev reflection in
clean semiconductor-superconductor interfaces and the new
possibilities such structures open, we hope to encourage experimental
work in this regime.

Our results were obtained using a semiclassical formalism, with which
we reduced a two-dimensional nonseparable problem to an effective 
one-dimensional problem.  We have demonstrated the usefulness of this
formalism in a few situations, and hopefully it can be used in the
future to solve other problems which are hard to tackle using the
conventional techniques.

We used this approach also for diffusive NS systems and demonstrated the
connection between the effects of reflectionless tunneling in
ballistic and diffusive NS junctions. 
We then considered the phenomena of reflectionless tunneling and the 
reduction in the density of states in diffusive NS junctions and 
showed that both can be obtained from a general criterion for the 
barrier, though large, to be transparent to pair current. 

\acknowledgments

We benefited from the valuable comments of V. Shumeiko. We gratefully 
acknowledge useful discussions with N. Argaman, N. M. Chtchelkatchev, 
M. H. Devoret, Y. M. Galperin, 
A. Krichevsky, R. de-Picciotto, A. Silva, and I. Ussishkin.
This work was supported by the Israel Academy of Science, by the 
German-Israeli Foundation (GIF), and by the Albert Einstein Minerva Center 
for Theoretical Physics at the Weizmann Institute.

\appendix

\section{Andreev reflection from an $N$ trajectory in SNS junctions}
\label{appSNS}

In this appendix we describe the recursion formalism leading to 
Eq.~(\ref{evenSNS}) and obtain a similar equation for trajectories 
that hit the NS interfaces an odd number of times. 

We choose the phase of the upper superconductor to be $\Phi_s/2$ and 
the phase of the lower superconductor to be $-\Phi_s/2$. The Andreev 
reflection amplitude of an electron hitting the upper (lower) boundary is 

\begin{equation} 
r_{he}^{+(-)}(1) = ire^{+(-)i\Phi_s/2} \; ,
\label{rhepm}
\end{equation}
and the Andreev reflection amplitude of an incoming hole is given by
$r_{eh}(1) = -r_{he}^*(1)$.  Throughout this appendix we consider
trajectories that hit the NS interfaces any number of times, with the
last hit occurring at the top interface. This is done for simplifying
the calculation, and since the final result is even in $\Phi_s$, it
does not depend on this assumption. However, the treatment for
trajectories that hit the interfaces an odd or an even number of
times has to be done separately.  For an odd trajectory, the
recursion relation is 

\begin{equation}
r_{he}(2N-1) = r_{he}^+(1) + 
\frac{r_{ee}^+(1) r_{he}(2N-2) r_{hh}^+(1)}{1-r_{eh}^+(1) r_{he}(2N-2)} \; , 
\label{oddrecursion}
\end{equation}
where $r_{ee(hh)}^+(1)$ is the normal reflection of an electron (hole)
from the upper NS interface. For an even trajectory the recursion
relation is 

\begin{equation}
r_{he}(2N) = r_{he}^-(1) + 
\frac{r_{ee}^-(1) r_{he}(2N-1) r_{hh}^-(1)}{1-r_{eh}^-(1) r_{he}(2N-1)} \; . 
\label{evenrecursion}
\end{equation}
We define 
\begin{equation}
r_{he}(2N-1) = r_{he}^-(1) \frac{\beta_{2N-1}}{\gamma_{2N-1}}
\label{bettagammaodd}
\end{equation}
and 
\begin{equation}
r_{he}(2N) = r_{he}^+(1) \frac{\beta_{2N}}{\gamma_{2N}} \; . 
\label{bettagammaeven}
\end{equation}
Using the relations written after Eq.~(\ref{rheN}), we obtain 

\begin{equation}
r_{he}(2N-1) = r_{he}^+(1) 
\frac{\beta_{2N-2}+\gamma_{2N-2}}{r^2\beta_{2N-2}+\gamma_{2N-2}} \; , 
\end{equation}
which we insert into Eq.~(\ref{evenrecursion}), and obtain the equation 

\begin{eqnarray}
\left( \begin{array}{c}\beta_{2N} \\ \gamma_{2N} \end{array} \right) & = & 
\left( \begin{array}{cc}1 + r^2 e^{-i \Phi_s} & 1+e^{-i \Phi_s}  
\\ r^2 (1+e^{i \Phi_s}) & 1 + r^2 e^{i \Phi_s}
\end{array} \right) 
\left( \begin{array}{c}\beta_{2N-2} \\ \gamma_{2N-2} \end{array} \right) 
\nonumber \\ & = & 
\left( \begin{array}{cc}1 + r^2 e^{-i \Phi_s} & 1+e^{-i \Phi_s}  
\\ r^2 (1+e^{i \Phi_s}) & 1 + r^2 e^{i \Phi_s}
\end{array} \right)^N 
\left( \begin{array}{c} 0 \\ 1 \end{array} \right) \, ,
\label{matrixSNS} 
\end{eqnarray} 
where the last equation is obtained by explicitly finding that 
\begin{equation}
\left( \begin{array}{c}\beta_2 \\ \gamma_2 \end{array} \right) =
\left( \begin{array}{c}1+e^{-i \Phi_s} \\ 1 + r^2 e^{i \Phi_s} 
\end{array} \right) \; . 
\label{twotwoSNS}
\end{equation} 
Following the same route for the odd case, we obtain 

\begin{eqnarray}
\left( \begin{array}{c}\beta_{2N+1} \\ \gamma_{2N+1} \end{array} \right) = 
\left( \begin{array}{cc}1 + r^2 e^{i \Phi_s} & 1+e^{i \Phi_s}  
\\ r^2 (1+e^{-i \Phi_s}) & 1 + r^2 e^{-i \Phi_s}
\end{array} \right) 
\left( \begin{array}{c}\beta_{2N-1} \\ \gamma_{2N-1} \end{array} \right) 
\nonumber \\ = 
\left( \begin{array}{cc}1 + r^2 e^{i \Phi_s} & 1+e^{i \Phi_s}  
\\ r^2 (1+e^{-i \Phi_s}) & 1 + r^2 e^{-i \Phi_s}
\end{array} \right)^N 
\left( \begin{array}{c} e^{i \Phi_s} \\ 1 \end{array} \right) \; . 
\label{matrixSNSodd} 
\end{eqnarray} 
Diagonalizing the matrices in Eqs.~(\ref{matrixSNS}) and (\ref{matrixSNSodd}) 
we obtain for the odd case 

\end{multicols}

\renewcommand{\theequation}{A\arabic{equation}}
\widetext

\noindent
\begin{picture}(3.375,0)
  \put(0,0){\line(1,0){3.375}}
  \put(3.375,0){\line(0,1){0.08}}
\end{picture}
\begin{equation}
R_{he}(2N+1,\Phi_s) = \left(z + (1-z) 
\coth^2{\left[y + 
N \tanh^{-1}\left({\frac{2r \cos(\Phi_s/2) \sqrt{(1-z)}} 
{1+r^2\cos{\Phi_s}}} \right) \right]} 
\right)^{-1} ,
\label{oddSNS} 
\end{equation} 
\hfill
\begin{picture}(3.375,0)
  \put(0,0){\line(1,0){3.375}}
  \put(0,0){\line(0,-1){0.08}}
\end{picture}

\begin{multicols}{2}
\noindent 
where 

\begin{equation}
y = \tanh^{-1}\left({\frac{r \sqrt{(1-z)}}
{e^{-i \Phi_s/2} + ir^2 \sin{(\Phi_s/2)}}}\right) \, ,
\end{equation}
and for the even case Eq.~(\ref{evenSNS}).  
These equations are similar, only in the latter $y=0$.

\renewcommand{\theequation}{B\arabic{equation}}
\renewcommand\thesection{\Alph{section}}

\section{Validity of the conductance formula for the diffusive slab}
\label{appfactor2}

In Sec.~\ref{secdiffusive} we obtain the linear conductance of a diffusive 
NIS junction using the approximation that the electron's motion in the 
normal slab is deterministic. However, for a normal slab with short range 
disorder Beenakker {\it et al.} \cite{BRM94} obtain a formula which differs 
in the $\Gamma \gg T, T \rightarrow 0$ limit by a factor of $2$ from 
Eq. (\ref{eq:Glimits}). 

In order to understand the factor of $2$ difference between the two cases 
we use the conductance formula for zero temperature of Beenakker {\it et al.}  
\cite{Bee92} 

\begin{equation} 
G_{NS}=\frac{4e^2}{\it{h}} \sum_{m=1}^{n} \frac{T_m^2}{(2-T_m)^2} , 
\label{eq:Been} 
\end{equation}
where $T_m$ is the $m$th transmission eigenvalue of the normal slab. 

The total transmission probability through the normal slab is given by $T =
\Sigma_{m=1}^{n} T_m$.  In our approximation of the long-range scattering
potential, each electron
entering the slab is predetermined, according to the position and
direction at the entrance, to be either transmitted through the slab
or reflected back to the normal reservoir (deterministic scattering). 
The transmission eigenvalues of the normal slab are therefore all zero 
and unity and in this case 
\begin{equation}
\sum_{m=1}^{n} \frac{T_m^2}{(2-T_m)^2} = \sum_{m=1}^{n} T_m = T  
\end{equation}
and 
\begin{equation}
G_{NS} = \frac{4e^2}{h} \sum_{m=1}^{n} T_m = \frac{4e^2}{h} T \; .  
\end{equation}
Since $G_N=(2e^2/h) \sum_{m=1}^{n} T_m$, we obtain the relation
$G_{NS} = 2G_N$.  However, in general, $T_m^2/(2-T_m)^2 < T_m$ for all
$0<T_m<1$.  In the case of short-range disorder the distribution of
the transmission eigenvalues is such that $\Sigma_{m=1}^{n}
T_m^2/(2-T_m)^2 = \frac{1}{2} \Sigma_{m=1}^{n} T_m = \frac{1}{2} T$ 
\cite{Bee92}, and therefore $G_{NS} = (2e^2/h) T = G_N$.  
This results in a factor of $2$ difference in the $T \rightarrow 0$
limit between our conductance formula (\ref{eq:Glimits}) and the
formula obtained by Beenakker {\it et al.} 

\renewcommand{\theequation}{C\arabic{equation}} 

\section{length of a diffusive N trajectory - random walk theory} 
\label{app1}
 
We are interested in the question of how long a trajectory in the normal 
metal has to be in order to hit the interface $N$ times. 
Looking at random walk in two dimensions on a lattice which is rotated 
by $45^\circ$ from the coordinate axes, it is easy to see that, since we are 
not interested at the exact point the trajectory hits the interface, there is 
a one-to-one correspondence between returning to the interface in two 
dimensions, and returning to the origin in the one-dimensional random walk.  
The question of return probabilities in one dimension is addressed in 
Ref.~\cite{Fel66}, Chap.~3, using random walk path theory. 
This approach is elementary and very instructive, and here we will just 
state its main results concerning our problem. 
Using path theory, Feller shows that in a one-dimensional random walk model,
the probability of a first return to the origin after $k$ steps is 
approximately,
for large $k$, $f_k=(2 \sqrt{\pi} k^{3/2})^{-1}$. Therefore, there is no
average length for the first return ($\sum_{k=0}^\infty k f_k$ diverges). This
peculiar result leads to a nonlinear dependence of the length of the
trajectory on the number of times it hits the interface. (If there were an
average return length $\alpha$, then the average length of a trajectory that
hits the interface $N$ times would be $\alpha N$).  It is further shown that
the probability to hit the interface $N$ times in a trajectory of length
smaller than $\hat{L}$ is a function of $\hat{L}/N^2 l \equiv w$, 
where $l$ is the mean free path and is given by 
\begin{equation}
P(w)=\sqrt{\frac{2}{\pi}} \int_{w^{-1/2}}^\infty e^{-s^2/2} ds \; .
\end{equation}
This means that in order to hit the interface $N$ times a particle has
to travel a length of order $N^2 l$.  A typical trajectory of length
$N^2 l$ that hits the interface $N$ times is not made of $N-1$ loops
of similar length.  The main contribution to the length of such a
trajectory comes from one or two of its longest loops, whose lengths are
of order $N^2 l$.  This arises from the fact that:
$\sum_{k=N^2}^\infty f_k \approx 1/N$, which means that if we have $N$
returns to the origin, about one of them is going to be longer than
$N^2 l$.  As $N$ increases, we have probability of order 1 to have a
loop of length $N^2 l$, and therefore the length of the longest loop,
as well as the length of the whole trajectory, is of order $N^2 l$.

\end{multicols}


\begin{thebibliography}{10}

\bibitem{KKG+91}
A. Kastalsky, A.~W. Kleinsasser, L.~H. Greene, R. Bhat, and F.~P. Milliken,
  Phys. Rev. Lett. {\bf 67},  3026  (1991).

\bibitem{NKH92}
C. Nguyen, H. Kroemer, and E.~L. Hu., Phys. Rev. Lett. {\bf 69},  2847  (1992).

\bibitem{WVMK92}
B.~J. van Wees, P. de~Vries, P. Magn\'{e}e, and T.~M. Klapwijk, Phys. Rev.
  Lett. {\bf 69},  510  (1992).

\bibitem{NATA92}
J. Nitta, T. Akazaki, H. Takayanagi, and K. Arai, Phys. Rev. B {\bf 46},  14286
   (1992).

\bibitem{TAN95}
H. Takayanagi, T. Akazaki, and J. Nitta, Phys. Rev. B {\bf 51},  1374  (1995).

\bibitem{TAN95b}
H. Takayanagi, T. Akazaki, and J. Nitta, Phys. Rev. Lett. {\bf 75},  3533
  (1995).

\bibitem{DHW+95}
A. Dimoulas, J.~P. Heida, B.~J. v.~Wees, T.~M. Klapwijk, W. v.~d. Graaf, and G.
  Borghs, Phys. Rev. Lett. {\bf 74},  602  (1995).

\bibitem{MHW+97}
A.~F. Morpurgo, S. Holl, B.~J. van Wees, T.~M. Klapwijk, and G. Borghs, Phys.
  Rev. Lett. {\bf 78},  2636  (1997).

\bibitem{MWK+97}
A.~F. Morpurgo, B.~J. van Wees, T.~M. Klapwijk, and G. Borghs, Phys. Rev. Lett.
  {\bf 79},  4010  (1997).

\bibitem{HWKB99}
J.~P. Heida, B.~J. van Wees, T.~M. Klapwijk, and G. Borghs, Phys. Rev. B {\bf
  60},  13135  (1999).

\bibitem{BKW83}
P.~A.~M. Benistant, H. van Kempen, and P. Wyder, Phys. Rev. Lett. {\bf 51},
  817  (1983).

\bibitem{BGKW85}
P.~A.~M. Benistant, A.~P. van Gelder, H. van Kempen, and P. Wyder, Phys. Rev. B
  {\bf 32},  3351  (1985).

\bibitem{SKHB00}
J.~H. Schon, C. Kloc, R.~C. Haddon, and B. Batlogg, Science {\bf 288},  656
  (2000).

\bibitem{SKB00}
J.~H. Schon, C. Kloc, and B. Batlogg, Nature {\bf 406},  702  (2000).

\bibitem{SL90}
E.~V. Sukhorukov and I.~B. Levinson, Sov. Phys. JETP {\bf 70},  782  (1990).

\bibitem{KZSJ95}
A. Kadigrobov, A. Zagoskin, R.~I. Shekhter, and M. Jonson, Phys. Rev. B {\bf
  52},  R8662  (1995).

\bibitem{CRL96}
N.~R. Claughton, R. Raimondi, and C.~J. Lambert, Phys. Rev. B {\bf 53},  9310
  (1996).

\bibitem{ASRL96}
N.~K. Allsopp, J.~S. Canizares, R. Raimondi, and C.~J. Lambert, J. Phys.
  Condens. Matter {\bf 8},  L377  (1996).

\bibitem{BKZ+98}
H.~A. Blom, A. Kadigrobov, A.~M. Zagoskin, R.~I. Shekhter, and M. Jonson, Phys.
  Rev. B {\bf 57},  9995  (1998).

\bibitem{Cht01}
N.~M. Chtchelkatchev, JETP Lett. {\bf 73},  94  (2001).

\bibitem{MB94}
A.~F. Morpurgo and F. Beltram, Phys. Rev. B {\bf 50},  1325  (1994).

\bibitem{Bee92}
C. Beenakker, Phys. Rev. B {\bf 46},  12841  (1992).

\bibitem{Sch00}
M. Schechter, {P}h.D. Thesis, The Weizmann Institute of Science, 2001.

\bibitem{WS96}
G. Wendin and V.~S. Shumeiko, Superlattices Microstruct. {\bf 20},  569
  (1996).

\bibitem{PADC93}
V.~T. Petrashov, V.~N. Antonov, P. Delsing, and R. Claeson, Phys. Rev. Lett.
  {\bf 70},  347  (1993).

\bibitem{PADC95}
V.~T. Petrashov, V.~N. Antonov, P. Delsing, and R. Claeson, Phys. Rev. Lett.
  {\bf 74},  5268  (1995).

\bibitem{AD96}
M.~P. Anantram and S. Datta, Phys. Rev. B {\bf 53},  16390  (1996).

\bibitem{TML01}
J. Torres, T. Martin, and G.~B. Lesovik, Phys. Rev. B {\bf 63},  134517
  (2001).

\bibitem{FLB98}
A.~L. Fauchere, G.~B. Lesovik, and G. Blatter, Phys. Rev. B {\bf 58},  11177
  (1998).

\bibitem{BH91}
C.~W. Beenakker and H. van Houten, Phys. Rev. B {\bf 43},  12066  (1991).

\bibitem{Argaman}
We thank Nathan Argaman and Doron Cohen for clarifying this point to us.

\bibitem{BRM94}
C.~W.~J. Beenakker, B. Rejaei, and J.~A. Melsen, Phys. Rev. Lett. {\bf 72},
  2470  (1994).

\bibitem{Vol94}
A. Volkov, Physica B {\bf 203},  267  (1994).

\bibitem{ZSZ95}
F. Zhou, B. Spivak, and A. Zyuzin, Phys. Rev. B {\bf 52},  4467  (1995).

\bibitem{Fel66}
W. Feller, {\em An Introduction to Probability Theory and Its Applications}
  (Wiley, New York, 1966).

\end{thebibliography}
\end{document}